\documentclass[12]{article}
\usepackage{graphicx}

\begin{document}

\title{SPECTROPOLARIMETRY AND INFRARED PHOTOMETRY
OF MAGNETIC WHITE DWARFS: VACUUM POLARIZATION EFFECT OR MAGNETIC CIA?}

\author {Gnedin\, Yu.N.$^1$, Borisov\, N.V.$^{2}$,
Larionov\, V.M.$^3$,\\Naidenov\, I.D.$^2$, Natsvlishvili\,
T.M.$^1$, Piotrovich\, M.Yu.$^1$}

\maketitle

\begin{center}
{\small (1) Central Astronomical Observatory at Pulkovo,
Saint-Petersburg, Russia.

(2) Special Astrophysical Observatory, Nizhnii Arhyz, Russia.

(3) Astronomical Institute of Saint-Petersburg State University,
Saint-Petersburg, Russia.}
\end{center}

\begin{abstract}
We present brief review of two probable physical mechanisms that
can explain the results of photometric and spectropolarimetric
observations of magnetic white dwarfs: vacuum polarization effect
into a strong magnetic field and, so-called, magnetic collision
induced absorption (magnetic CIA). Both mechanisms provide
observed rotation of polarization ellipse and suppression of
spectral energy distributions. The results of spectropolarimetric
observations of magnetic white dwarfs made at Russian BTA-6m and
the results of the near infrared photometric observations with
Russian-Italian AZT-24 telescope located at Campo Imperatore are
also presented.
\end{abstract}

\section{Introduction}

We present the results of spectropolarimetric and infrared
photometric observations of a number of isolated magnetic white
dwarfs. The spectropolarimetric observations were made at russian
6-m telescope (BTA-6m). The IR photometry of magnetic white dwarfs
were made at AZT-24 telescope of Central Astronomical Observatory
at Pulkovo, that is now installed at Campo Imperatore, Italy. The
IR observations were made in frame reference of a program of
Pulkovo, Rome and Teramo Observatories. We found deficiency of IR
fluxes of some magnetic white dwarfs: GrW+70.8247, G99-47,
WD1658+441, G240-72, GD229.

Spectropolarimetric observations of GrW+70.8247 and GD229 by
BTA-6m showed the rotation of polarization ellipse, the
polarization ellipse of GrW+70.8247 being rotated by $90^{0}$. We
analyzed the optical and NIR continuum polarization of five
magnetic white dwarf stars observed by West, 1989. We consider
both effects of the polarization ellipse rotation and suppression
of infrared fluxes for highly magnetized dwarfs as an indirect
evidence of vacuum polarization effect though the collision
induced absorption (CIA) into magnetized atmosphere can also
explain effect of infrared fluxes suppression.

\section{Short review of vacuum polarization effect by a magnetic
field and its astrophysical manifestations}

The high magnitudes of magnetic fields of neutron stars and white
dwarfs give rise to new effects in the traditional physical
processes involving interaction of radiation with matter. One of
most important effects is so-called polarization of
electron-positron plasma by a strong magnetic field. Just as in an
ordinary magnetoactive plasma the photon propagation in a
magnetized vacuum is also described in terms of two normal modes
(waves) with different polarization states and refractive indices
$n_{1,2}$. The polarization of the vacuum itself is due to virtual
$e^+ e^-$ pairs and becomes significant when the magnetic field
strength

\begin{equation}
B\geq B_C=\frac{m_e^2 c^3}{e\hbar}=4.414\times 10^{13} G \label{1}
\end{equation}

\noindent where $B_C$ is the magnetic field value at which the
electron cyclotron energy $\hbar\omega_B=\frac{\hbar eB}{m_e c}$
is equal to electron rest mass energy $m_e c^2$. Nevertheless it
appeared that the vacuum polarization must be taken into account
in the analysis of many radiation processes even if the magnetic
strength $B\ll B_C$.

\subsection{}

In his excellent review Adler, 1971, presented the expressions for
refractive indices of normal modes in the magnetized vacuum at $B
< B_C$ and $\hbar\omega < m_e c^2$:

\begin{equation}
n_1=1+\frac{7}{90\pi}\frac{e^2}{\hbar
c}\left(\frac{B_\bot}{B_C}\right)^2;\,\,\,\,\,
n_2=1+\frac{2}{45\pi}\frac{e^2}{\hbar
c}\left(\frac{B_\bot}{B_C}\right)^2 \label{2}
\end{equation}

\noindent where $B_\bot=B\sin{\theta}$ and $\theta$ is the angle
between the photon wave vector and the magnetic field directions.

The normal modes in this case are linearly polarized, the electric
vector of mode 1 oscillating in the magnetic field and wave vector
plane and that of mode 2 oscillating in the perpendicular plane.
Vacuum polarization effect modifies the dielectric property of the
medium and the polarization of photon modes propagated in a
magnetoactive plasma, thereby altering the radiative scattering
and absorption opacities (see Pavlov and Gnedin, 1984, and
Meszaros, 1992, for reviews).

The existence of quite strong magnetic fields of neutron stars and
white dwarfs provides possibilities to search the vacuum
polarization effects in astrophysical observations of compact
objects. Novick et al, 1977, were the first who have considered
the possibility to measure of the phase shift between the vacuum
polarization modes in radiation of neutron stars. Pavlov and
Gnedin, 1984, were the first who mentioned the importance of the
vacuum polarization effect and for magnetic white dwarfs. The
following step was to analyze the interaction of radiation with a
"mixture" of vacuum and plasma in a strong magnetic field (Gnedin
et al., 1978, Pavlov and Gnedin, 1984). The modern detailed
analysis of this situation was made in the series of papers by Lai
and Ho, 2002, 2003, Ho and Lai, 2001, 2003, 2004, Ho et al., 2003.

\subsection{}

The first important step for estimation of the vacuum polarization
effect is to calculate the magnitude of the phase shift $\varphi$
between the two normal waves due to the difference in their phase
velocities:

\begin{equation}
\varphi=\frac{\omega}{c}\int|n_1-n_2|dl=\frac{l}{5\times
10^{-7}cm}\frac{\hbar\omega}{m_e
c^2}\left(\frac{B_\bot}{B_C}\right)^2
\label{3}
\end{equation}
For neutron stars (NS) at $\hbar\omega=1KeV$, $B_\bot=4\times
10^{12}G$ the magnitude $\varphi\sim 1$ after transversing a very
small $l=0.3mm\ll R_{NS}$ distance ($R_{NS}$ is a radius of NS).
It means that the radiation of NS will be partially depolarized
via the so-called Cotton-Mouton effect, which is the analog to the
familiar Faraday effect for a medium in which the normal modes are
polarized linearly. In this case the polarization ellipse
"oscillates" around the direction of polarizations of the normal
waves, changing the ratio of the axes and the direction of
rotation of the electric vector in an oscillatory manner. It may
lead to a perfect depolarization of circularly polarized radiation
from a NS and to partially depolarized the linearly polarized
radiation (except the cases when the electric vector lies in the
${\bf K B}$ plane or at right angle to it).

For WDs:

\begin{equation}
\varphi=1.2\left(\frac{\hbar\omega}{3eV}\right)\left(\frac{B_\bot}{4\times
10^8 G }\right)^2\left(\frac{R_{WD}}{10^9 cm} \right) \label{4}
\end{equation}

\noindent in the optical spectral range. The situation for WDs
looks better because of there is no complete depolarization in
this situation. In a result the possibility arises to search the
vacuum polarization effect in the optical spectral range via the
polarimetric observations.

\subsection{}

Especially interesting effects arise if one analyzes the
interaction of radiation with a "mixture" of vacuum and plasma in
a strong magnetic field due to the different types of the
anisotropy in plasma and vacuum. These effects arise in the region
where the contribution from the vacuum to polarization of normal
waves is of the same order of magnitude as that from the plasma.
Specifically there are two values of photon energy at which the
contributions of the vacuum and plasma on the linear polarization
of normal modes cancel out each other. This case is called by
"vacuum resonance". One of these specified energies lies in the
region of cyclotron energy $\hbar\omega_B$ and corresponds to the
vacuum resonance number density

\begin{equation}
N_{V,1}=\frac{1}{60\pi^2}\left(\frac{m_e
c}{\hbar}\right)^3\left(\frac{\hbar\omega_B}{m_e
c^2}\right)^4\cong 3\times 10^8\left(\frac{B}{4\times 10^8
G}\right)^4 cm^{-3} \label{5}
\end{equation}

Another "vacuum resonance" phenomenon can be existed in the region
outside the cyclotron energy if only the vacuum resonance number
density is to be:

\[
N_{V,2}=6\times
10^{19}Y_e^{-1}\left(\frac{E}{1KeV}\right)^2\left(\frac{B}{10^{12}}\right)^2
cm^{-3}\,\,\,for\,\,NS
\]
\begin{equation}
N_{V,2}=10^8 Y_e^{-1}\left(\frac{1\mu
m}{\lambda}\right)^2\left(\frac{B}{3\times 10^8}\right)^2
cm^{-3}\,\,\,for\,\,WD \label{6}
\end{equation}

\noindent where $Y_e$ is an electron fraction. In the completely
ionized plasma $Y_e=\frac{Z}{A}$.

The location of the vacuum resonance photon (wavelength) at a
given number density is:

\[
NS:\,\,\, E_V=0.24\left(\frac{Y_e N_V}{6\times
10^{19}}\right)^{1/2}\left(\frac{10^{12}}{B}\right)KeV
\]
\begin{equation}
WD:\,\,\, \lambda_V=0.283\left(\frac{10^8}{Y_e N_V
}\right)^{1/2}\left(\frac{B}{3\times 10^8}\right)\mu m
\end{equation}

Neutron stars and white dwarfs are characterized by different
situation. For neutron stars the vacuum resonance lies into the
deep layers atmosphere (photosphere) of a star (Lai and Ho, 2002,
2003, Ho and Lai, 2002, Ho et al., 2003, Potekhin and Chabrier,
2003, 2004). For magnetic WDs the number density value $\leq 10^8
cm^{-3}$ lies only in the most upper layer of the atmosphere
($N_V\sim 10^8 cm^{-3}$ corresponds to the distance $l\sim 20H$,
where $H$ is the density scale height if only the electron
fraction is not extremely low) or into the plasma environment
(coronas or plasma envelopes produced by the pressure of cyclotron
radiation, see Zheleznyakov, 1997, Bespalov and Zheleznyakov,
1990, Zheleznyakov and Serber, 1991). Namely Zheleznyakov and his
colleagues showed that the pressure of cyclotron radiation in the
magnetic WD photosphere can be compared and even can surpass the
gravity force. Then hydrostatic equilibrium of plasma on magnetic
white dwarfs can be disrupted by large radiation pressure and the
radiation-driven ejection from the white dwarf photosphere can be
possible. Zheleznyakov and his colleagues called this situation
"radiation discon" object (see fig.8,9 from Zheleznyakov, 1997,
book). They claimed that the structure of plasma envelopes of
magnetic WDs with the effective temperature $T_e\geq 10^4 K$ is
drastically different from the structure of thin hot corona. If
the plasma density of such an envelope is large enough, it can
strongly distort the photosphere spectrum and give rise to the
broad and deep depressions bands in the observed radiation
spectrum.

Also one needs to take into consideration that the influence of
strong large scale magnetic fields on the structure and
temperature distribution in WD atmospheres. For example, Fendt and
Dravins, 2000, displayed that magnetic fields may provide an
additional component of pressure support, thus inflating the
atmosphere compared to non-magnetic case. They found
quantitatively that a mean surface poloidal field strength 100 MG
and a toroidal field strength of 10 MG may increase the scale
height at least by factor 10.

\subsection{}

Let us now consider the basic effects arising if photons are
propagating across the vacuum resonance. The first main effect is
changing the orientation of the polarization ellipse. It can
rotate by the definite angle $\leq 90^0$. The magnitude of the
rotation angle is dependent on the peculiarities of the plasma
region at the vacuum resonance because the orthogonality of normal
modes in the resonance region may be violated. The rotation of the
polarization ellipse is result of resonant conversion of photon
modes across the vacuum resonance (Gnedin and Pavlov, 1984, Lai
and Ho, 2002, 2003).

Lai and Ho, 2002, investigated this process in detail and showed
that the physics of this mode conversion is analogous to the
Mikheyev-Smirnov-Wolfenstein mechanism for neutrino oscillations.
They have demonstrated that the conversion process is more
effective if the adiabatic condition is fulfilled at resonance.
The last one requires for MWD:

\begin{equation}
E_{con}\geq 1.5eV\left(\frac{10^9 cm}{R_{WD}}\right) \label{8}
\end{equation}

In this case the adiabatic probability of conversion is
$P_{con}=1-\exp{(-\frac{\pi}{2}\frac{E}{E_{con}})}$. The jump
probability can be calculated with the Landau-Zener formula:
$P_j=\exp{(-\frac{\pi}{2}\frac{E}{E_{con}})}$ (Lai and Ho, 2002,
2003). This process is accompanied without the essential
conversion of photon modes.

The second important for observations effect is the suppression of
Rayleigh-Jeans region of the black body spectrum and, partially,
the proton cyclotron lines for neutron stars and other spectral
lines (Ho and Lai, 2003). For magnetic WDs (MWDs) the essential
modification of the electron cyclotron lines is realized because
in the "vacuum+plasma" mixture the ordinary wave acquires also
cyclotron resonance and increases the cyclotron absorption (Pavlov
and Gnedin, 1984, Zheleznyakov, 1997). In the Zheleznyakov
radiation-driven discon model of MWD the increase of cyclotron
absorption may strongly distort the photospheric spectrum and give
rise to the broad and deep depression bands in the observed
radiation from such radiation-driven discon (Zheleznyakov, 1997).

In conclusion of this section one can say that the vacuum
polarization may produce the observable effects in the radiation
from radiation-driven discon of a magnetic white dwarf.

\section{Magnetic collision induced absorption by Rydberg states into
magnetic white dwarf}

Here we suggest the complete analogy to the vacuum polarization
effect that may act in partially ionized atmospheres of MWDs. Our
main idea consists of the fact that in partially ionized plasma
also it may exist the resonance region where contribution to the
dielectric constant from non- and ionized components may cancel
out each other. Even in the non-magnetized plasma such situation
may arise because the refractive index of this plasma is equal:
$n=1+2\pi N_H \alpha_H - \frac{\omega_p^2}{2\omega^2}$ where $N_H$
is the density of a neutral component, $\alpha_H$ is the
polarizability of a single atom (molecule), $\omega_p$ is electron
plasma frequency.

For hydrogen non-magnetized plasma the resonance energy is
$E_R\approx 10 eV \sqrt{\frac{N_e}{N_H}}$. In the strong magnetic
fields of WDs and NSs the atoms, especially in their high excited
states acquires non-spherical shape and may be oriented by a
strong magnetic field.

Therefore we introduce, pure formally, for magnetized non-ionized
plasma:

\begin{equation}
n_1=1+2\pi N_a\alpha_{||};\,\,\,\, n_2=1+2\pi N_a\alpha_\bot
\label{9}
\end{equation}

\noindent where the polarizability $\alpha_{||}$ corresponds to
the case when the electric vector of the electromagnetic wave lies
in the (${\bf K B}$) plane, $\alpha_\bot$ corresponds the electric
vector orientation perpendicularly to the (${\bf K B}$) plane.

Let us consider the case when $\alpha_{||} > \alpha_\bot$. This
case is namely realized in a strong magnetic field. Atomic
structure is affected by strong magnetic fields. It is well-known
(see, for example, the book by Dolginov et al., 1995), that the
critical value of the field at which the essential reform of an
atom becomes important is reached if the cyclotron energy
$\hbar\omega_B$ is compared to the Rydberg energy. This condition
implies a field strength:

\begin{equation}
B>B_0=\frac{Z^2 m_e^2 e ^3 c}{\hbar^3}=2.35 Z^2 \times 10^9 G
\label{10}
\end{equation}

If $B\gg B_0$, the magnetic forces acting on an electron of an
atom dominate over the Coulomb forces, the transverse size of the
atom becoming less than the Bohr radius and the transverse
velocity of the electron becoming greater than its longitudinal
velocity.

The Eq.(10) means that the Bohr radius of an hydrogen atom
$r_0=\frac{\hbar^2}{m_e e^2}$ becomes larger that the so-called
the magnetic length $a_m=(c\hbar / eB)^{1/2}$ that is namely
determined the transverse size of an atom. The atom acquires the
ellipsoidal cigar shape instead of the typical spherically
symmetric form.

The magnetic field strength (10) is rather high for the typical
magnetic white dwarfs. Therefore the neutron stars are namely
suitable targets for the investigation of the behavior of atoms
and molecules in a strong magnetic fields (Dolginov et al., 1995,
Potekhin and Pavlov, 1997, Potekhin and Chabrier, 2003, 2004, Ho
et al., 2003).

However there can exist the situation when atoms become
anisotropic and in the magnetic white dwarfs with the typical
magnetic field strengths $B\sim 10^6 \div 10^8 G$. Such situation
is really existed if the atoms appear in strongly excited
(Rydberg) states. For an atom in highly $n\gg 1$ exciting state
its characteristic size is $r_n\approx r_0 n^2$.

For example, Bethe and Salpeter, 1957, give for the average radius
of highly excited state:

\begin{equation}
\langle r_n^3\rangle = \frac{n^2}{8Z^3}[21n^4 + 35n^2 +
4]\,\,\overrightarrow{n\longrightarrow\infty}\,\,\frac{21}{8Z^3}n^6
\label{11}
\end{equation}

or for a hydrogen atom:$\langle
r_n^3\rangle^{1/3}=(21/8)^{1/3}n^2$

Thus for an strongly excited atom the critical magnetic field
strength (10) is

\begin{equation}
B_0\cong \frac{1.7}{n^2}\times 10^9 G
\label{12}
\end{equation}

For the magnetic whit dwarf GrW+70.8247 the pole magnetic field
strength $B_p=3.2\times 10^8 G$ and this value is the critical one
if the hydrogen atom is found in the excited state with $n>3$.

Now it is possible to get the total analogy of magnetic CIA to the
vacuum polarization effect. Let us consider the case
$\alpha_{||}>\alpha_\bot$ and suggest for the total analogy to the
vacuum polarization: $\alpha_{||}/\alpha_\bot=7/4$.

Following to the analogy between birefringences of magnetized
vacuum and magnetized highly excited atomic states, one can obtain
from Eqs.(2) and (9) the relation:

\begin{equation}
\left(\frac{B_\bot}{B_C}\right)^2\equiv \frac{180\pi^2\hbar c}{7
e^2}\alpha_{||} N_a
\label{13}
\end{equation}

In mixed hydrogen and helium gases, colliding pairs of atoms and
molecules such as $H_2-H_2$, $H_2-He$, $H-He$ can be sources of,
so-called, Collision Induced Absorption (CIA) opacity. Another
source of CIA opacity is the origin of highly excited (Rydberg)
states of atoms in the photosphere of a magnetized white dwarf,
via also the collision process. The originated Rydberg states
acquire a large dipole moment and therefore can produce strong
absorption in infrared range of spectrum.

Collision induced infrared absorption was discovered by Welsh and
associates in 1949 in an attempt to observe an infrared absorption
band of oxygen dimers. In pure and mixed gases a great variety of
collision induced absorption spectra is now known. Several review
of experimental work have been published (see, for instance,
Welsh, 1972, van Kranendonk, 1980, Borysow and Frommhold, 1985,
and also the 1981 October issue of the Canadian Journal Physics
which contained a special section devoted to collision induced
phenomena).

The process of CIA have been intensively applicable to stellar
atmospheres, especially to white dwarf atmospheres (Bergeron et
al., 1995, Borysow et al., 1997, Rohrmann et al., 2002). Recent
detail calculations of $H_2-He$ absorption coefficients are
available from Jorgensen et al., 2000. Most up-to-date CIA
opacities due to $H_2-H_2$ have been calculated by Borysow et al.,
2001, for $1000<T<7000K$ and frequencies between $20-20000
cm^{-1}$. $H-He$ CIA data are now available for temperatures
$1500-10000 K$ and frequencies $50-11000 cm^{-1}$.

Chemi-ionization and chemi-recombination processes in low
temperature layers of white dwarf atmospheres are very important
for producing helium atom Rydberg states population in weakly
ionized layers of helium-rich DB white dwarfs (see Mihajlov et
al., 2003). But all these processes (CIA, chemi-ionization and
chemi-recombination) have been considered only for non-magnetized
white dwarf atmospheres.

We present here only the phenomenology of the process producing
atoms Rydberg states population taking into account the anisotropy
of highly excited atomic states into strong magnetic fields of
magnetized white dwarfs (see Eqs.(9)-(13)). The presence of a
magnetic field can increase highly excited Rydberg atoms.
Remarkably the magnetic field induces a permanent electric dipole
moment of the atom (Lesanovsky et al., 2003, Raithel et al.,
1993). Raithel et al., 1993, found in the experiments the evidence
for large permanent electric dipole moments of Rydberg atoms in
crossed electric and magnetic fields. They found that the dipole
moments have a large value if the scaled electric field strength
have the value $EB^{-4/3}=0.75$ (with electric and magnetic field
strengths in atomic units). In the atmosphere of a white dwarf an
atom is exposed to the electric fields of surrounding atoms, ions
and free changes. The motional Stark effect gives an electric
field perpendicular to the magnetic one. The mean values of the
electric field felt by each atom in the atmosphere of a magnetic
white dwarf can be $\geq 10^8 V/m$, i.e. these values correspond
to the scaling relation of Raithel et al., 1993. It is also very
important that a mean surface poloidal field strength of $\sim 100
MG$ and a toroidal field strength of $2-10 MG$ can increase a
scale height by a factor of $\geq 10$ (Fendt and Dravins, 2000).

Let us estimate the resonance number density for magnetic CIA
(MCIA) in the atmosphere of a magnetic white dwarf:

\begin{equation}
N_V\cong 10^{18} Y_e^{-1}(E/3eV)\left(\frac{180\pi^2\hbar c }{7
e^2}\alpha_{||} N_a\right)=10^{18}Y_e^{-1}(E/3eV)(3.5\times
10^4\alpha_{||} N_a)
\label{14}
\end{equation}

Let us estimate the number density $N_a$ of Rydberg state atoms
required for the resonance number density $N_V\geq 10^{18}
cm^{-3}$ at the level of white dwarf atmosphere:

\begin{equation}
N_a\geq \frac{7}{180\pi^2}\frac{e^2}{\hbar
c}\frac{1}{\alpha{||}}\approx 4\times
10^{19}\left(\frac{\alpha_H}{\alpha_{||}}\right) \label{15}
\end{equation}

Here $\alpha_H=0.67\times 10^{-24} cm^{-3}$ is the classical
polarizability of a free hydrogen atom without a magnetic field.
The polarizabilities of highly excited atoms and quasimolecules
are radically increased with the main quantum number:
$\alpha_{||}\approx \alpha_{H}n^6$. For $n=10$ the required number
density $N_a\geq 4\times10^{13} cm^{-3}$, i.e. at $\sim$six
magnitude lesser the typical number density in a white dwarf
atmosphere. For $n=3$ $N_a\geq 10^{16} cm^{-3}$.

Correspondingly, the expression for the location of resonance
photon energy takes a form:

\begin{equation}
E_V=3\left(\frac{Y_e N_V}{10^{18}}\right)\left(\frac{180\pi^2\hbar
c}{7 e^2}\alpha_{||} N_a\right)^{-1}=3\left(\frac{Y_e
N_V}{10^{18}}\right)\left(\frac{\alpha_{||}}{\alpha_H}\right)
\left(\frac{N_a}{4\times 10^{19}}\right)
\label{16}
\end{equation}

The adiabatic condition is to be:

\begin{equation}
E_{ad}>7.6\left(\frac{1 cm}{H}\right)^{1/3}eV\approx 0.35 eV
\label{17}
\end{equation}

\noindent where $H=kT_{WD}R_{WD}^2/GM_{WD}m_p$ is the height of
homogeneous atmosphere of a white dwarf. Its value for the
magnetic dwarf GrW+70.8247 is $\sim 10^4 cm$ ($T_{WD}\approx 10^4
K$, $R_{WD}\approx 10^9 cm$, $M_{WD}\approx 0.5 M_\odot$).

\section{The results of spectropolarimetric and infrared
photometric observations of magnetic white dwarfs}

The basic observational phenomena of both physical effects
considered above and connected with the fact vacuum polarization
or existence of highly excited Rydberg states of atoms into a
strong magnetic field modify the dielectric properties of a medium
and the polarization modes, altering the radiative scattering and
absorption opacities.

The basic effects across resonance are

(a) Change the orientation of the polarization ellipse by $\leq
90^0$ without the helicity changing.

(b) Suppression of Rayleigh-Jeans region of black body spectrum
and cyclotron spectral lines. Both effects have been observable,
including our spectropolarimetric observations by BTA-6m and
infrared photometric observations by AZT-24 telescope in Campo
Imperatore, Italy.

Still at 1989 West demonstrated measurements of the near infrared
polarization of GrW+70.8247 combined with the optical
spectropolarimetry of Landstreet and Angel, 1975. The near
infrared polarimetry over the region $1.1-1.64 \mu m$ was obtained
with IRPOL (West et al., 1988) attached to the MMT between 1985
July and 1986 November. West results demonstrated the real
rotation of the orientation of the polarization ellipse exactly by
$90^0$ near the wavelength $\lambda =0.6 \mu m$.

Fig.1 presents the results of our spectropolarimetric observations
of GrW+70.8247 at 1999 July. The spectropolarimetric observations
display the gradual rotation of the polarization ellipse at the
wavelength range $\lambda\lambda 4800-5000$ \AA. This transition
can be considered as a result of adiabatic conversion from the
low-opacity X-mode to the high-opacity O-mode (and via versa) as
it crosses the region where contributions of polarizabilities from
usual magnetoactive plasma and vacuum (or Rydberg atomic states)
compensate each other. One ought to take into consideration the
existence of the absorbing feature at the region near $\lambda =
4150$ \AA. This feature exists in all Stokes parameters but if the
linear polarization $P_l$ looks as depression, the circular
polarization V displays the excess at the level $P_V\sim -5\%$.
One of the possible explanation of this feature is the "vacuum
resonance" lying in the region of cyclotron energy (see Eq.(5)).

At 2002 we repeated the spectropolarimetric observations of
GrW+70.8247. These results will be presented at next publications.

Fig.2 displays the wavelength dependence of Stokes parameters for
the magnetic white dwarf GD229 obtained in the result of our
observations at BTA-6m. This figure demonstrates also the jump of
rotation angle of the polarization ellipse in the spectral region
$\lambda\lambda$ 4200-4600 \AA.

This behavior of the position angle reveals probably the
phenomenon of magnetoelectric Jones birefrigence and dichroism
(Jones effect) (Graham and Raab, 1983). This phenomenon reveals
that certain uniaxial media may exhibit birefrigence and dichroism
along axes which are at $\pm 45^{\circ}$ relative to the axis of
anisotropy (the magnetic field direction in our case). In this
case the additional contribution appears to the position angle
when the direction of light propagation is perpendicular to the
magnetic field:

\begin{equation}
\Delta \varphi_J=\frac{2\pi}{\lambda}\int
[n_{+45}(\lambda)-n_{-45}(\lambda)]dl
\label{18}
\end{equation}

Recently Budker and Stalnaker, 2003, suggested that the
interference between atomic magnetic dipole and electric field
induced dipole transition amplitudes provides magnetoelectric
Jones effect.

These spectropolarimetric observations (Fig.2) have been made at
BTA-6m SAO RAN in 2002 July with the long slit spectrograph UAGS
and the analyzer of the linear and circular polarization (Naidenov
et al., 2002) located at the main focus of the telescope. The
detector was CCD-camera with 1024x1024 pixels. The size of a pixel
is 24x24 $\mu m$. The resulting dispersion was a 2.4 \AA /pixel,
in the spectral range $\lambda\lambda 3500-8000$ \AA. The seeing
was 1".2. It allows to get the resulting spectral resolution $\sim
5$ \AA. For reprocessing data the system MIDAS has been used.
HD204827 has been used as spectropolarimetric standard (Turnshek
et al., 1990).

Near IR observations of magnetic white dwarfs were obtained at the
AZT-24 1.1m telescope in Campo Imperatore (Italy) with SWIRCAM
during a period spanning July-August 2003. SWIRCAM is the infrared
camera that incorporates 256x256 HgCdTe NIGMOS 3-class (PICNIC)
detector at the focus of AZT-24. It yields a scale of 1".04/pixel
resulting in a field of view 4x4 sq. arcmin. The observations were
performed through standard JHK broadband filters. The list of our
targets includes GrW+70.8247, GD229, G240-72, GD356, G227-35,
WD1031-234, WD1312+098.

The spectral energy distribution (SED) for GrW+70.8247 is
presented at Fig.3. Here the Black Body SED is also presented. We
see the real fact of SED suppression in JHK spectral ranges.
Another example of such suppression of G99-37 SED is presented at
Fig.4 (We used here only known observed data, see Bergeron et al.,
2001).

\section{Conclusions: distinctions between vacuum polarization
and magnetic CIA effects}

The polarimetric jump of the orientation of the polarization
ellipse by the angle $\leq 90^0$ and the suppression of SED in
near infrared region of spectra of white dwarfs with a strong
magnetic field display two possible physical mechanisms
responsible for both phenomena: vacuum polarization or magnetic
CIA. The critical difference between these both mechanisms is the
characteristic scale factor. For the vacuum polarization this
scale factor is $l=R_S$, for the magnetic CIA it is
$l=H=kTR_S^2/GM_{WD}m_{H}$, i.e. the homogeneous atmosphere
height. It means that vacuum polarization effect displays
existence an extended region like an extended corona around a
white dwarf. The most suitable physical situation is the discon
model of Zheleznyakov and his co-authors (e.g. Zheleznyakov, 1997,
and refs. in). It was shown that the structure of plasma envelopes
of magnetic white dwarf is drastically different. The pressure
force by photospheric radiation at cyclotron frequencies can
exceed the gravitational force acting on a proton and can display
ejection of plasma from the photosphere. In a result the formation
of an extended envelope in the white dwarf magnetosphere as well
as the disk near the magnetic equator are produced. Zheleznyakov,
1997, called this phenomenon "radiative-driven discon". If the
plasma density of such an envelope is quite high ($N_e\geq 10^8
cm^{-3}$), it may distort the photospheric spectrum and produce
the broad and deep depression bands in the observed radiation from
magnetic white dwarf. The discon-like structure can also display
the rotation of the orientation of the polarization vector if one
takes into account the effect of vacuum resonance mentioned above.

The some physical situation can be originated in the result of CIA
process in the magnetized photosphere of a white dwarf with a
strong magnetic field. Strong magnetic field produces the
orientation of photospheric atoms. CIA process of oriented atoms
and molecules into magnetized photosphere displays simultaneously
the rotation of orientation of polarization vector and suppression
of the spectral energy distribution of a magnetic white dwarf.
This physical situation does not require the existence of extended
magnetosphere of a white dwarf. The basic difficulty for such
regime is quite fast ionization of Rydberg atoms embedded in a
photospheric plasma of a magnetized white dwarf (e.g. Vanhaecke et
al., 2004).

We suggest more detail analysis of both physical regimes in the
next paper.

\section*{Acknowledgements}

This work is partially supported by RFBR Grant \# 03-02-17223, the
Program of the Presidium of RAN "Nonstationary Phenomena in
Astronomy" and by the Program of Astronomy of Russian Science and
Education Ministry.

\begin{figure*}
\includegraphics[height=22cm]{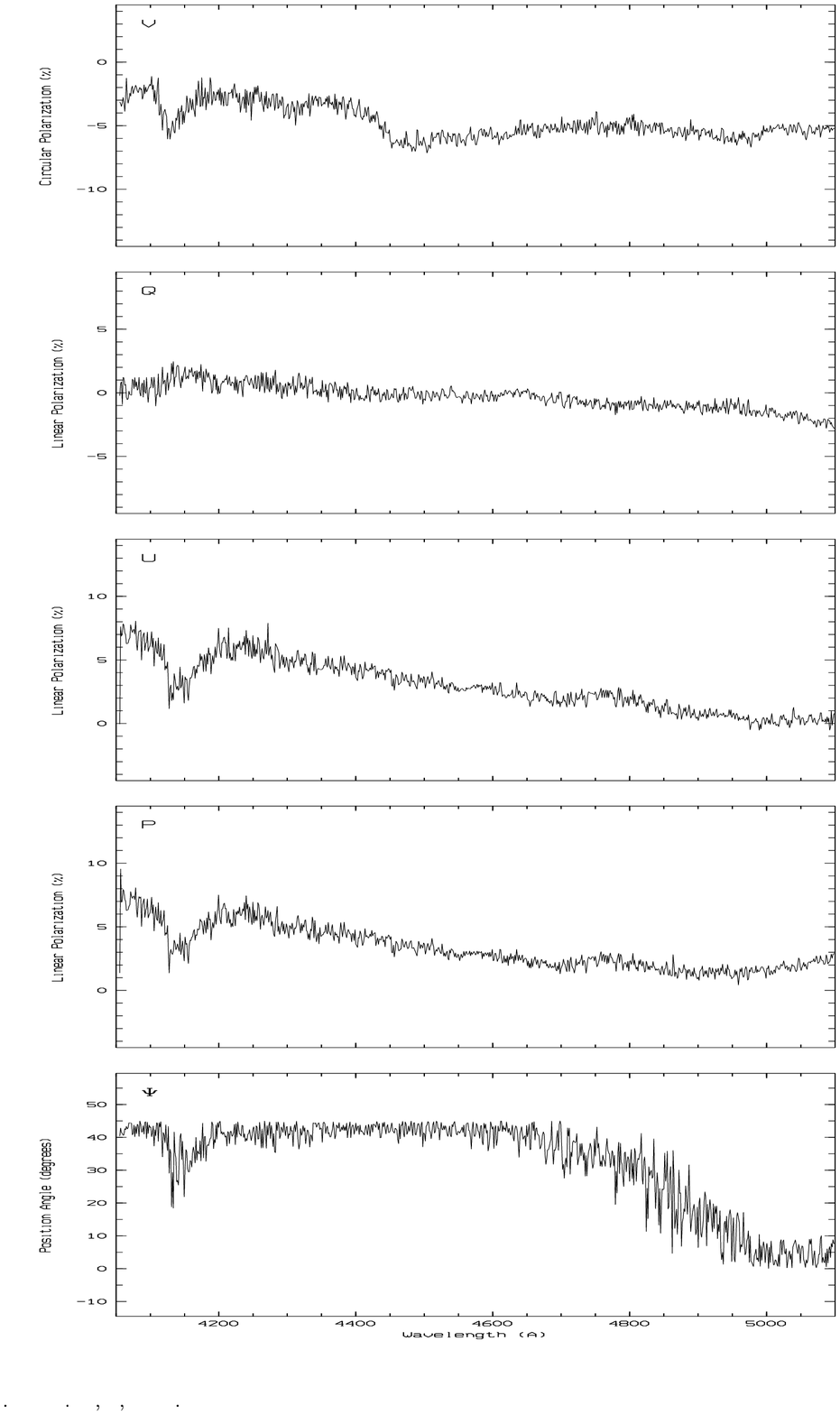}
\caption{The result of polarimetric observations of the magnetic
white dwarf GrW+70.8247. The wavelength dependence of the Stokes
parameters (from the top): V, Q, U, the degree of the linear
polarization P and the positional angle $\psi$.}
\end{figure*}

\begin{figure*}
\includegraphics[width=16cm]{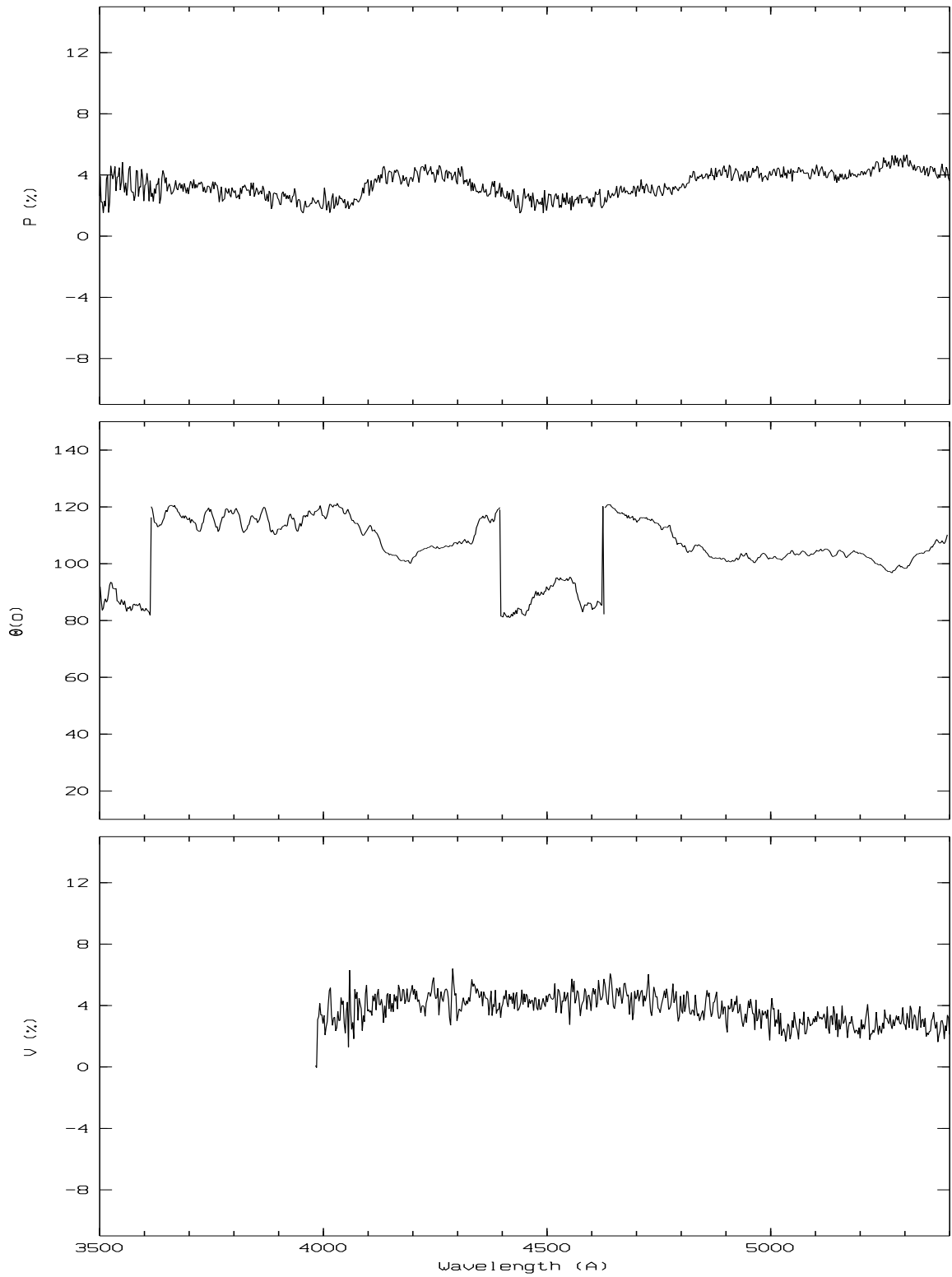}
\caption{The results of polarimetric observations of the magnetic
white dwarf G229. The wavelength dependence of the linear
polarization degree P, the position angle $\theta$ and of the
circular polarization V.}
\end{figure*}

\begin{figure*}
\includegraphics[width=16cm]{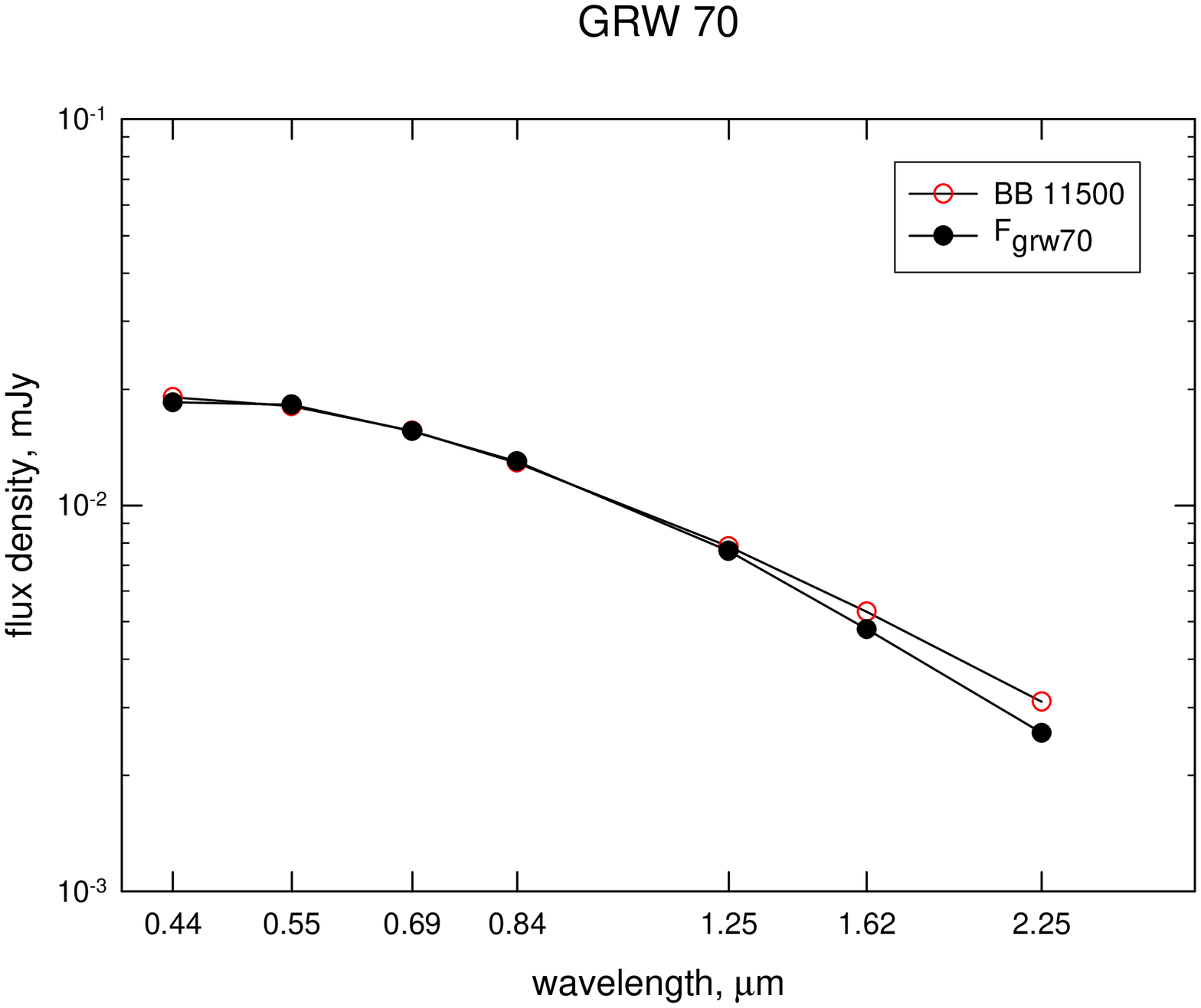}
\caption{The spectral energy distribution (SED) of the magnetic
white dwarf GrW+70.8247 (black circles). The black body
distribution is shown by open circles.}
\end{figure*}

\begin{figure*}
\includegraphics[width=16cm]{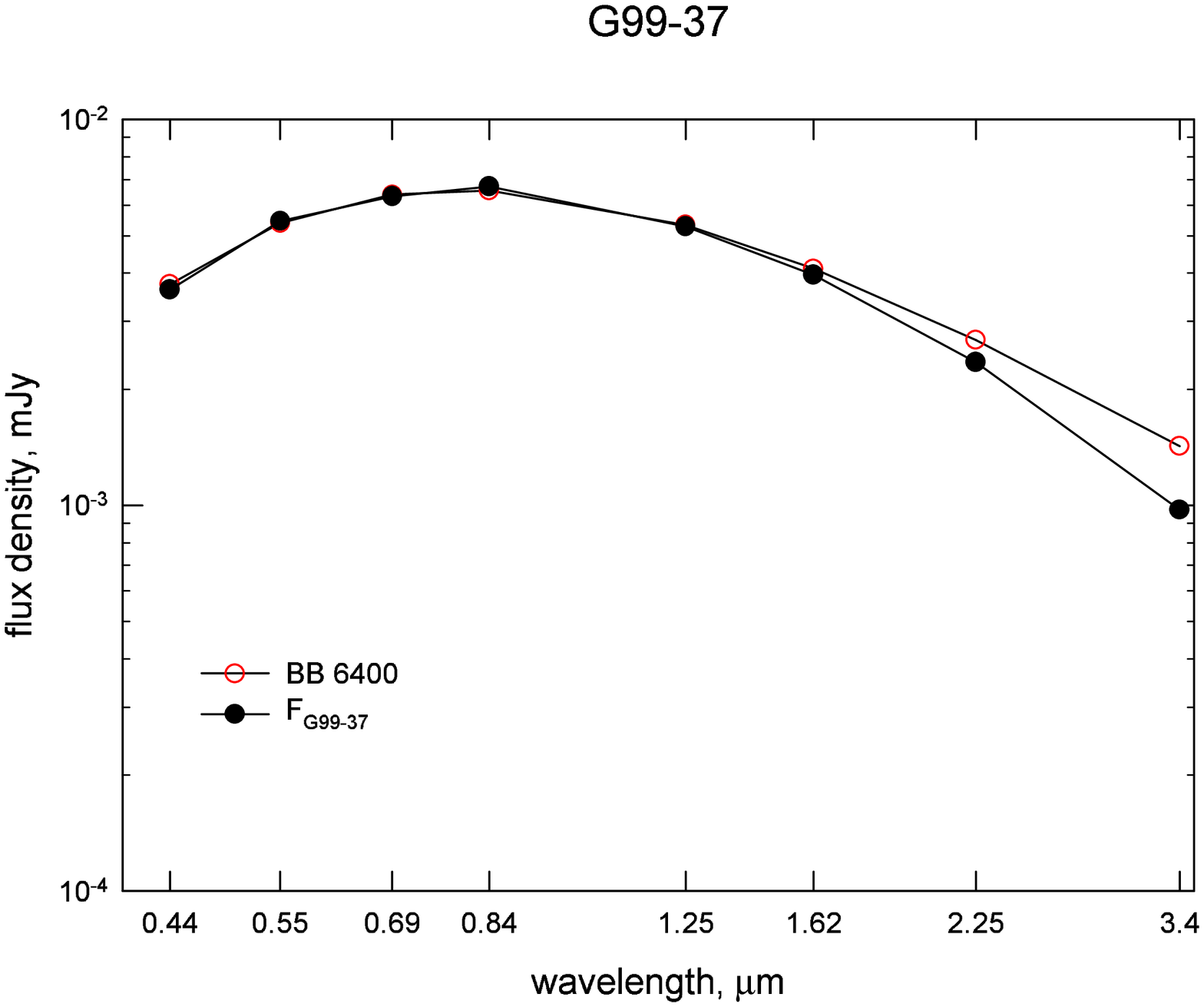}
\caption{The same distributions as at Fig.3 for the magnetic white
dwarf G99-37.}
\end{figure*}


\begin{thebibliography}{99}
\bibitem{1}Adler S. L., Annals. of Phys. (N.Y.), {\bf 67}, 599 (1971).
\bibitem{2}Bergeron P., Wesemael F., Beauchamp A., PASP, {\bf 107}, 1047 (1995).
\bibitem{3}Bespalov P. A., Zheleznyakov V. V., Pis'ma Astron. Zh., {\bf 16}, 1030 (1990).
\bibitem{4}Bethe H. A., Salpeter E. E., {\it Quantum Mechanics of One and Two-electron atoms} (Academic Press, New York, 1957).
\bibitem{5}Borysow A., Jorgensen U. G., Zheng C., Astron.Astrophys., {\bf 324}, 185 (1997).
\bibitem{6}Borysow J., Frommhold L., in Proc. Of the NATO Advanced Research Workshop. (Premium Press, 1985).
\bibitem{7}Budker D., Stalnaker J. E., physics/0302096 (2003).
\bibitem{8}Dolginov A. Z., Gnedin Yu. N., Silant'ev N. A., {\it Propagation and Polarization of Radiation in Cosmic Media} (Gordon and Breach Publ., Bazel., Switzerland, 1995).
\bibitem{9}Fendt C., Dravins D., Astron.Nachr., {\bf 3}, 193 (2000).
\bibitem{10}Gnedin Yu. N., Pavlov G. G., Shibanov Yu. A., Sov.Astron.Lett., {\bf 4}, 117 (1978).
\bibitem{11}Graham E. B., Raab R. E., Proc.R.Soc.Lond. {\bf A390}, 73 (1983).
\bibitem{12}Ho W. C. G., Lai D., Ap.J., {\bf 607}, 420 (2004).
\bibitem{13}Ho W. C. G., Lai D., MNRAS, {\bf 327}, 1081 (2001).
\bibitem{14}Ho W. C. G., Lai D., MNRAS, {\bf 338}, 233 (2003).
\bibitem{15}Ho W. C. G., Lai D., Potekhin A. Y., Chabrier G., Ap.J., {\bf 599}, 1293 (2003).
\bibitem{16}Jorgensen U. G., Hammer D., Borysow A., Falkesgaard J., Astron.Astrophys., {\bf 361}, 283 (2000).
\bibitem{17}Lai D., Ho W. C. G., Ap.J., {\bf 566}, 373 (2002).
\bibitem{18}Lai D., Ho W. C. G., Ap.J., {\bf 588}, 962 (2003).
\bibitem{19}Landstreet J. D., Angel J. R. P., Ap.J., {\bf 196}, 819 (1975).
\bibitem{20}Lesanovsky I., Shmiedmayer J., Schmelcher P., Europhys.Lett., physics/0312045 (2003).
\bibitem{21}Meszaros P., {\it High-Energy Radiation from Magnetized Neutron Stars} (Chicago: Univ. Chicago Press, 1992).
\bibitem{22}Mihajlov A. A., Ignjatovic Lj. M., Dimitrijevic M. S., Djuric Z., Ap.J.S.S., {\bf 147}, 369 (2003).
\bibitem{23}Naidenov I. D., Valyavin G. G., Fabrika S. N., Borisov N. V., Burenkov A. N., Vikul'ev N. A., Moiseev S. V., Kudryavtsev D. O., Bychkov V. D., Bull.SAO, {\bf 53}, 124 (2002).
\bibitem{24}Novick R. M., Weisskopf M. C., Angel J. R. P., Sutherland P. G., Astrophys.J.Lett., {\bf 215}, L117 (1977).
\bibitem{25}Pavlov G. G., Gnedin Yu. N., Astrophys.Sp.Phys., {\bf 3}, 197 (1984).
\bibitem{26}Potekhin A. Y., Chabrier G., Ap.J., {\bf 585}, 955 (2003).
\bibitem{27}Potekhin A. Y., Chabrier G., Ap.J., {\bf 600}, 317 (2004).
\bibitem{28}Potekhin A. Y., Pavlov G. G., Ap.J., {\bf 483}, 414 (1997).
\bibitem{29}Raithel G., Fauth M., Walther H., Phys.Rev., {\bf A47}, 419 (1993).
\bibitem{30}Rohrmann R. D. et al., MNRAS, {\bf 335}, 499 (2002).
\bibitem{31}Rohrmann R. D., Serenelli R .D., Althaus L. G., Benvenuto O. C., astro-ph/0205084 (2002).
\bibitem{32}Turnshek D. A., Bohlin R. C., Williamson R. L. et al., Ap.J., {\bf 99}, 1243 (1990).
\bibitem{33}van Kranendonk J., Editor {\it Intermolecular Spectroscopy and Dynamical Properties of Dense Systems} (Soc.Italiana di Fisica, Bologna, 1980).
\bibitem{34}Vanhaecke N., Comparat D., Duncan A., Tate A., Pillet P., quant-ph/0401045 (2004).
\bibitem{35}Welsh H. L., MTP Int.Rev.Sci., "Spectroscopy", Phys.Chem., Ser.1, {\bf 3}, A. D. Buckingham and D. A. Ramsey Eds., Butterworth, London (1972).
\bibitem{36}West S. C., Ap.J., {\bf 345}, 511 (1989).
\bibitem{37}West S. C., Schmidt G. D., Pawlicki R., Rieke G. H., Angel J. R. P., Rudy R. J., PASP, {\bf 100}, 859 (1988).
\bibitem{38}Zheleznyakov V. V., {\it Radiation in Astrophysical Plasma} (1997).
\bibitem{39}Zheleznyakov V. V., Serber A. B., Pis'ma Astron. Zh., {\bf 17}, 419 (1991).
\end{thebibliography}
\end{document}